# Matryoshka Phonon Twinning in α-GaN

Bin Wei[1,2,3], Qingan Cai[3], Qiyang Sun[3], Yaokun Su[4], Ayman H. Said[5], Douglas L. Abernathy[6], Jiawang Hong[1]*, and Chen Li[3,4]*

[1]School of Aerospace Engineering, Beijing Institute of Technology, Beijing 100081, China
[2]Henan Key Laboratory of Materials on Deep-Earth Engineering, School of Materials Science and Engineering, Henan Polytechnic University, Jiaozuo 454000, China.
[3]Department of Mechanical Engineering, University of California, Riverside, Riverside, CA 92521, USA2
[4]Materials Science and Engineering, University of California, Riverside, Riverside, CA 92521, USA
[5]Advanced Photon Source, Argonne National Laboratory, Lemont, IL 60439, USA
[6]Neutron Scattering Division, Oak Ridge National Laboratory, Oak Ridge, TN 37831, USA

*Correspondence: chenli@ucr.edu; hongjw@bit.edu.cn

**ABSTRACT**

Understanding lattice dynamics is crucial for effective thermal management in high-power electronic devices because phonons dominate thermal transport in most semiconductors. This study utilizes complementary inelastic X-ray and neutron scattering techniques and reports the temperature-dependent phonon dynamics of α-GaN, one of the most important third-generation power semiconductors. A prominent Matryoshka phonon dispersion is discovered with the scattering tools and confirmed by the first-principles calculations. Such Matryoshka twinning throughout the three-dimension reciprocal space is demonstrated to amplify the anharmonicity of the related phonon modes through creating abundant three-phonon scattering channels and cutting the phonon lifetime of affected modes by more than 50%. Such phonon topology effectively contributes to the reduction of the in-plane thermal transport, thus the anisotropic thermal conductivity of α-GaN. The results not only have significant implications for engineering the thermal performance and other phonon-related properties of α-GaN, but also offer valuable insights on the role of anomalous phonon topology in thermal transport of other technically important semiconductors.

## 1. Introduction

Gallium nitride (GaN), one of the most important third-generation power semiconductors, excels in power density and high-temperature stability due to its wide bandgap and high thermal conductivity [1-4] (230 W m$^{-1}$ K$^{-1}$ at room temperature has been reported [5-8]), among many other favorable properties. To further miniaturize high-power electronics [4], great efforts have been devoted to studying the thermodynamics of $\alpha$-GaN (wurtzite structure, see Fig. S1a in Supplementary Information). However, the knowledge on its phonon dynamics remains limited, with key questions yet to be answered. For example, existing experimental measurements of the phonon dispersion relation of $\alpha$-GaN are limited to ambient conditions, and the phonon scattering processes and the temperature effects are mostly unexplored [9, 10]. Moreover, the anisotropic thermal transport of $\alpha$-GaN along *a* (in-plane) and *c* (out-of-plane) axis directions remains controversial due to the challenges in transport measurements [7,11-18] and calculations [19-21]. Therefore, a more comprehensive understanding of the phonon dynamics is pivotal to investigating the thermal transport and other thermodynamics properties of GaN.

Engineering of phonon topology is one of the main strategies to manipulate thermal properties through phonon engineering besides doping [22], creating solid solution [23], isotopic engineering [24, 25] and nanostructuring [26, 27] Many novel properties, such as low thermal conductivity [28-30], negative thermal expansion [31], and anomalous phase transitions [32], can be attributed to exotic phonon dispersion topology, e.g., crossing/anti-crossing behaviors [28, 29] bunched acoustic phonon, and dispersion waterfall [33]. Recently, local phonon dispersion nesting behaviors have been shown to augment acoustic-optical three-phonon scattering channels, amplify anharmonicity, and suppress lattice thermal transport [34, 35].

Here we report a novel in-plane Matryoshka-like phonon dispersion twinning, in which the optical and the acoustic phonon dispersions are like nesting dolls (see Fig. S2). Such behavior is observed in $\alpha$-GaN single crystals by both inelastic X-ray scattering (IXS) and inelastic neutron scattering (INS), and is confirmed by first-principles calculations. The Matryoshka phonon twinning provides a great magnitude of acoustic-optical scattering channels and contributes to the reduction of the in-plane thermal conductivity, leading to the anisotropic thermal conductivity of $\alpha$-GaN. This result is supported by the phonon lifetime measured through scattering linewidths and provides valuable insights into phonon topology engineering for thermal management.

## 2. Materials and Methods

### 2.1 GaN Single Crystal Samples:

High-quality GaN single crystals (un-doped, n-type, MTI Corporation [36]) used in this work were grown by the hydride vapor phase epitaxy (HVPE)-based method with low dislocation density ($< 1 \times 10^7$ cm$^{-2}$) (Fig. S1c). The quality of the crystals was checked with both X-ray and neutron diffraction. The full width at half maximum (FWHM) of the X-ray diffraction peak at (002) plane is around $0.10 \pm 0.02°$ (Fig. S1d).

### 2.2 Inelastic X-ray Scattering Measurements:

High-resolution IXS experiment was performed to measure the phonon dispersions of an $\alpha$-GaN single crystal (250 μm thickness, the lower panel of Fig. S1c) at 50, 175, and 300 K. The measurements were conducted at 30-ID-C (the High-resolution Inelastic X-ray Scattering beamline, HERIX) at the Advanced Photon Source (APS) [37, 38]. The incident photon energy was ~23.7 keV with an energy resolution $\Delta E$ of 1.2 meV and a momentum resolution of 0.65 nm$^{-1}$. The single crystal was attached to a copper post by varnish and the copper post was mounted in a closed-cycle cryostat. The IXS measurements were accomplished at constant wave-vector mode in reflection geometry. The orientation matrix was defined by using Bragg peaks at (4 0 0), (0 0 4), (0 0 5), and (2 2 0) respectively.

**2.3 Inelastic Neutron Scattering Measurements:**

The INS measurements were carried out on a bigger single crystal ($\varphi$=5 cm, upper panel of Fig. S1c) on the Wide Angular-Range Chopper Spectrometer (ARCS), time-of-flight (TOF) neutron spectrometer at the Spallation Neutron Source (SNS) at Oak Ridge National Laboratory [39]. An incident energy of $E_i$ = 50 meV and a Fermi chopper frequency of 420 Hz were used to optimize instrument energy resolution. At the elastic scattering, the energy resolution is 2 meV. For INS, the [110] axis was set vertical, and $H\bar{H}K$ scattering plane was selected to obtain both the in-plane and out-plane phonons at 14, 50, 300, and 630 K. When collecting the data, 1° step was used with a rotation range from -90° to 90° for 300 K, and 2° step was used with a rotation range from -70° to 50° for other temperatures. Data reduction was performed using the Mantid program [40]. The data were normalized by the accumulated incident neutron flux, and the detector efficiency correction was applied based on the incoherent scattering of the vanadium standard.

**2.4 Scattering Data Processing:**

INS data shown in the present work were corrected for the sample temperature to obtain the dynamical susceptibility, $\chi''(\mathbf{Q}, E) = S(\mathbf{Q}, E)/[n_T(E)+1]$, where $n_T(E)$ stands for the Bose distribution, $S(\mathbf{Q}, E)$ is the four-dimensional scattering dynamical structure factor [41]

$$S(\mathbf{Q}, E) \propto \sum_s \sum_\tau \frac{1}{\omega_s} \left| \sum_d \frac{\bar{b}_d}{\sqrt{M_d}} \exp(-W_d) \exp(i\mathbf{Q} \cdot \mathbf{r}_d)(\mathbf{Q} \cdot \mathbf{e}_{ds}) \right|^2$$
$$\times \langle n_s+1 \rangle \delta(\omega-\omega_s) \delta(\mathbf{Q}-\mathbf{q}-\tau) \quad (1)$$

where $\bar{b}_d$ is the neutron scattering length, $\boldsymbol{Q} = \boldsymbol{k} - \boldsymbol{k'}$ is the wave vector transfer, and $\boldsymbol{k'}$ and $\boldsymbol{k}$ are the final and incident wave vector of the scattered particle. $\boldsymbol{q}$ is the phonon wave vector, $\omega_s$ is the eigenvalue of the phonon corresponding to the branch index $s$, $\tau$ is the reciprocal lattice vector, $d$ is the atom index in the unit cell, $\boldsymbol{r}_d$ is the atom position, $W_d$ is the corresponding Debye-Waller factor, and $\boldsymbol{e}_{ds}$ is the phonon eigenvectors. For given $\boldsymbol{Q}$, the measured scattering spectra were fitted with a damped-harmonic-oscillator (DHO) model [42]

$$S(\omega) = [1 + n(\omega)] \frac{1}{\pi M} \frac{\gamma \omega}{(\omega - \omega_0)^2 + (\gamma \omega)^2} \quad (2)$$

where $M$ is the effective mass, $\omega_0$ is the bare phonon energy in the absence of damping forces, and $2\gamma$ is a damping factor that describes the phonon scattering rates. $n$ is the Bose distribution. The bare phonon energy and phonon linewidths were extracted by de-convoluting with both instrument energy and momentum resolution functions, the latter of which eliminates the slope effect of highly dispersive phonons on linewidths.

**2.5 Simulation Methods:**

The first-principles calculations were performed based on the density functional theory (DFT) as implemented in the Vienna Ab Initio Simulation Package (VASP) [43]. The exchange−correlation energy was computed using the local-density approximation (LDA) functional, and the projector-augmented-wave (PAW) potentials were used ($4s^24p^1$ for Ga and $2s^22p^3$ for N). For plane-wave expansion in reciprocal space, we used 600 eV as the kinetic energy cutoff value. The $\boldsymbol{q}$-mesh was chosen as 7×7×5. The convergence criteria for total energy was set to $1 \times 10^{-8}$ eV, and that for atomic force was $10^{-3}$ eV/Å. The structure was fully relaxed and the lattice constants, $a = b = 3.188$ Å, $c = 5.190$ Å, are slightly smaller than the experimental values ($a = b = 3.191$ Å, $c = 5.191$ Å obtained from HERIX at 300 K). The phonon dispersion of *α*-GaN at the level of harmonic approximation was calculated using the Phonopy package [44] with 4 × 4 × 3 supercell of *α*-GaN containing 192 atoms.

## 3. Results and discussion

### 3.1 Matryoshka phonon twinning behavior in *α*-GaN

The phonon dispersions of *α*-GaN were measured by IXS below the phonon gap around 45 meV, as shown in Fig. 1 for 300 K (the dispersion measurements at 50 and 175 K can be found in Fig. S3). The measurements are in excellent agreement with first-principles calculations. Most importantly, two parallel sections were found along Γ–M and Γ–K directions in the Brillouin zone (BZ), showing dispersion nesting behavior [34]: the near-constant energy difference is around 16 meV along Γ–M direction while it is around 15 meV along Γ–K direction. The upper nested branch consists of low energy optical (LEO) phonons in the lower $q$ range and longitudinal acoustic (LA) phonons in the higher $q$ range (referred to as the "Arc" branch in the following text), while the lower nested branch is one of the transverse acoustic branches (TA$_2$). Figs. 1b-1e show how the TA$_2$ and the Arc branches track each other with momentum transfer at 300 K (Other phonon branch evolutions and the temperature dependence of some optical phonons are shown respectively in Figs. S4 and S5). It can be clearly observed that, along both Γ–M and Γ–K directions, the energies of these phonon modes exhibit similar dispersive behaviors with increasing $q$, leading to a nearly parallel section between the branches.

An INS experiment was performed to measure the full phonon lattice dynamics in the reciprocal space to complement the IXS data, which only cover selected high symmetry directions. The acquired dynamical susceptibility, $\chi''(\mathbf{Q}, E)$, is shown in Fig. 2 as two-dimension slices along Γ–M, Γ–K, and Γ–A directions at 14 and 300 K, overlaid with the first-principles phonon dispersion calculation. The $\chi''(\mathbf{Q}, E)$ calculations (Figs. 2c, 2f, and 2i) based on first-principle phonon eigenvectors agree well with the INS measurements (Figs. 2b, 2e, and 2h). The nesting behavior of the TA$_2$ and the Arc branches can be clearly observed along Γ–M direction in both the experimental and the calculated $\chi''(\mathbf{Q}, E)$. The nesting behavior of the TA$_2$ and the Arc branches can also be observed along Γ–K, although the intensity of $\chi''(\mathbf{Q}, E)$ is weaker due to the scattering structure factor.

Jointly analyzing the IXS and INS results, it is remarkable that the phonon dispersion nesting behavior is exhibited throughout the basal plane in the BZ of α-GaN (Fig. S1b), as visualized by the volume and cross-section views of the measured $\chi''(\mathbf{Q}, E)$ at 300 K in Figs. 3a-3d (the results at 14 K are shown in Fig. S6). Similar to the phonons along the high symmetry directions in Fig. 2, the phonons along the non-high symmetry directions exhibit similar nesting behavior (additional cuts and visual angles can be found in Fig. S7). Such phenomenon can be vividly illustrated by the phonon dispersion surfaces in Fig. 3e from the first-principles calculation (also shown by the calculated phonon dispersion in Fig. S8a). In the basal plane of the reciprocal space, the nested TA$_2$ and Arc phonon dispersion surfaces are similarly cone-shaped and nest together in a Matryoshka-like twinning behavior (Fig. 3e). Such Matryoshka twinning is reflected by both the first-principles calculations and INS projection contours (Figs. 3f and 3g and Fig. S8b). Such Matryoshka twinning may provide vast three-phonon scattering channels in α-GaN by reducing the momentum transfer constraint, thus expanding the phonon scattering phase space. For example, in a three-phonon emission process where one phonon mode on the Arc branch ($\boldsymbol{q}_{Arc}$, $\omega_{Arc}$) decays into two phonon modes ($\boldsymbol{q}_1$, $\omega_1$) and ($\boldsymbol{q}_2$, $\omega_2$), for the mode ($\boldsymbol{q}_1$, $\omega_1$) on the Arc branch near Γ point, we can always find a TA$_2$ mode ($\boldsymbol{q}_2=\boldsymbol{q}_{Arc} - \boldsymbol{q}_1$, $\omega_1=\omega_{Arc} - \omega_1$) in the basal plane that enables such scattering (Fig. S9). This scattering behavior is consistent with recent report of weighted phonon scattering space in α-GaN, which shows the emission process is significantly larger than that of the absorption process in the energy range from 20 to 40 meV (~5 THz to ~10

THz), (see Fig. S10) [6] in line with the energy range of the Arch branch in the Matryoshka twinning.

**3.2 The impact of Matryoshka phonon twinning on the anharmonic scattering rate**

Phonon linewidths of $\alpha$-GaN at 300 K were extracted from the IXS measurements by deconvoluting both the instrument energy and momentum resolution functions, the latter of which is equally important due to the steep phonon dispersion. Phonon linewidth is proportional to the total scattering rate from all processes [34, 45] and dominated by anharmonic phonon-phonon interactions in $\alpha$-GaN. As shown in Fig. 4a, while the linewidths of most phonon modes are between 0.5 and 1.5 meV, the linewidths of the phonon modes on the $TA_2$ and the Arc branches are much broader and double of some other phonon modes. These large linewidths indicate that the nested phonons are scattered more strongly than the others, leading to the phonon lifetime of nested modes is cut by more than 50%. This result can also be observed in the INS data (Figs. 4b and 4c), in which the linewidths of phonon modes on the $TA_2$ and Arc branches are much larger than those on the high energy optical (HEO) branch along $\Gamma$–M direction and those in the HEO and LEO branches along $\Gamma$–A direction. The INS measurement is consistent with the Raman spectroscopy [46] result that the linewidth of the Arc branch at the $\Gamma$ point is 0.8 meV (Fig. 4d). The anharmonic phonon scattering rate is expressed as [47]

$$\tau_{qs}^{-1} = 2\Gamma_{qs} = \frac{36\pi}{\hbar^2} \Sigma_{q_1,q_2,s_1,s_2} |V(\mathbf{q},\mathbf{q}_1,\mathbf{q}_2,s,s_1,s_2)|^2$$
$$\times (n_1+n_2+1)\left[\delta(\omega_1+\omega_2-\omega_{qs})-\delta(\omega_1+\omega_2+\omega_{qs})\right]$$
$$+(n_2-n_1)\left[\delta(\omega_1-\omega_2-\omega_{qs})-\delta(\omega_1-\omega_2+\omega_{qs})\right] \quad (3)$$

where $V$ is the third-order interatomic potential, $\omega_{qs}$ is the phonon frequency of the phonon mode of wavevector $\mathbf{q}$ and branch index $s$, and $n$ is Bose occupation. This anharmonic scattering rate is determined by both the third-order interatomic potentials and the phonon scattering phase space: the former represents the scattering strength, and the latter represents the phase space of available scattering channels [48]. The large linewidths of the phonons involved in the Matryoshka twinning suggest that such behavior significantly promotes three-phonon scatterings in $\alpha$-GaN.

To elucidate the anharmonicity of the phonon modes on the Arc branch in $\alpha$-GaN, the evolution of the phonon energy at $q = 0.1$ along Γ–M direction was fitted over a wide temperature range with the following expression [49]

$$\omega_A(T) = \omega(0) - A(1 + 2/(e^{\frac{\hbar\omega(0)}{2k_BT}} - 1)) \tag{4}$$

where A and $\omega(0)$ are fitting constants, and $\omega_A(T)$ is the temperature-dependent phonon energy. The result in Fig. 5a shows a moderate anharmonicity below the Debye temperature (~636 K) [9], supported by the moderate difference between experimental isobaric [50] and calculated volumetric mode Grüneisen parameters (Fig. 5b), see the details in Supplementary Information. Additionally, the frozen phonon potential for the phonon mode on the Arc branch at $q = 0.1$ along Γ–M direction (Fig. 5c) only slightly departs from the harmonic behavior, indicating moderate anharmonicity (the phonon eigenvectors are shown in the insert in Fig. 5c). Considering such moderate anharmonicity, it is confirmed that the large linewidths of the phonon modes in the $TA_2$ and Arc branches are dominated by the enlarged scattering phase space from the vast scattering channels induced by the Matryoshka twinning.

## 3.3 The impact of Matryoshka phonon twinning on the anisotropic thermal conductivity

The anisotropic thermal conductivity of $\alpha$-GaN recently attracted intensive attention because it is critical for heat management in power electronics, where $\alpha$-GaN is a key component. However, the answer remains controversial: some reported that the out-of-plane thermal conductivity is greater than the in-plane one while others showed opposite results [5, 6, 19, 20] Recent reports suggest that the in-plane thermal conductivity is smaller than the out-of-plane one, with a relatively strong anisotropy of 12% [19, 20]. Here, we will investigate the roles of phonon group velocity and scattering rates on the anisotropic lattice thermal conductivity.

The temperature-dependent acoustic phonon group velocities were extracted from the measured phonon spectra (Fig. 2), as shown in Table 1. The acoustic phonon group velocities along the out-of-plane (Γ–A) direction are slightly larger than those along the in-plane (Γ–M and Γ–K) directions. The ratios of acoustic phonon group velocities at the same temperature are used to quantify the group velocity anisotropy. Along Γ–M directions, the acoustic phonon group velocities are 1%, 1.7%, and 0.8% larger for the LA branch, and 3.7%, 3%, and 4% larger for the $TA_2$ branch, at 14, 50, and 300 K, respectively. This behavior indicates that the anisotropy of acoustic phonon group velocity is minor and may not induce a sizable anisotropic thermal conductivity in $\alpha$-GaN.

The effects of in-plane Matryoshka twinning on lattice thermal conductivity is more significant in $\alpha$-GaN, as shown by the anomalously large linewidths of the phonon modes involved. The Matryoshka twinning amplifies the acoustic-optical three-phonon scattering in $\alpha$-GaN and reduces the in-plane thermal conductivity, especially for temperatures above the Debye temperature, where the lattice thermal conductivity is dominated by the phonon-phonon interactions. Therefore, the anisotropy of thermal conductivity is further enhanced by this twinning behavior in $\alpha$-GaN. If the Matryoshka phonon twinning could be suppressed by phonon engineering through strain or doping, the in-plane thermal transport properties of $\alpha$-GaN may potentially be enhanced for better thermal management at high temperatures.

## 4. Conclusion

Through a combination of inelastic X-ray/neutron scattering experiments and first-principles calculations, a novel Matryoshka phonon twinning is found throughout the basal plane of the Brillouin zone in $\alpha$-GaN. This phenomenon creates a vast number of three-phonon scattering channels and leads to enhanced anharmonic phonon scattering, shown by anomalously large phonon linewidths. The strong in-plane phonon scatterings due to the Matryoshka phonon twinning suppress the in-plane thermal conductivity of $\alpha$-GaN and enhance the thermal transport anisotropy, while its acoustic phonon group velocities remain fairly isotropic. Amplification and suppression of such Matryoshka phonon twinning by phonon engineering could provide a valuable means to control lattice thermal transport in many related materials, such as electronic devices, thermoelectrics, and thermal barriers.

**Conflict of interest**

The authors declare no conflict of interest.

**Acknowledgements**

The work at Beijing Institute of Technology is supported by the National Natural Science Foundation of China with Grant No. 11572040 and Beijing Natural Science Foundation (Grant No. Z190011). This work is supported by University of California, Riverside via Initial Complement. Theoretical calculations were performed using resources of the High-Performance Computing Center in University of California, Riverside. B.W. thanks the Joint PhD Program of Beijing Institute of Technology. Inelastic neutron scattering measurements used resource at the Spallation Neutron Source, a Department of Energy (DOE) Office of Science User Facility operated by the Oak Ridge National Laboratory. Inelastic X-ray scattering measurements used resource at Advanced Photon Source (APS), a Department of Energy (DOE) Office of Science User Facility operated by Argonne National Laboratory (ANL).

**Author Contributions**

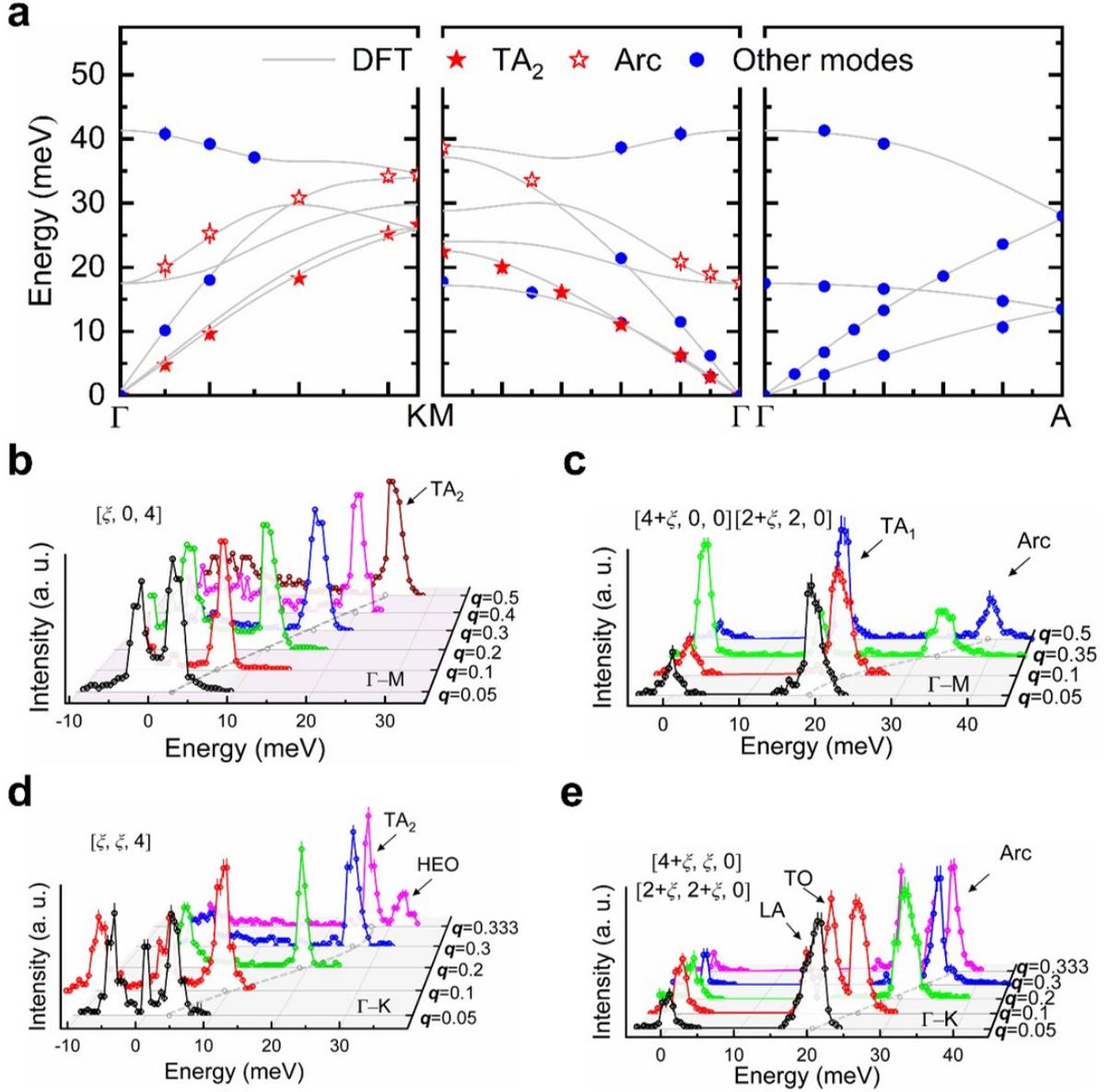

**Fig. 1. Phonon dispersion nesting observed by IXS measurements in $\alpha$-GaN.** (**a**), The measured phonon dispersion of $\alpha$-GaN by IXS at 300 K overlays on the calculation (grey lines). Along Γ−M and Γ−K, the TA$_2$ (solid stars) and the Arc (hollow stars) jointly show the nesting behavior. The energy of Arc branch at Γ point was measured by Raman spectroscopy. The error bars are from fitting uncertainties and comparable to or smaller than the data point symbols. (**b**)-(**e**), The measured phonon spectra evolution of TA$_2$ and Arc branches with increasing *q* along Γ−M and Γ−K at 300 K (reciprocal lattice units—r.l.u.). The phonon mode trends are reflected by the grey dash line with the peak centers labeled by circles. TA$_2$ branches were measured in the (004) zone along scattering wavevector **Q** = [$\xi$, 0, 4] (Γ−M) and **Q** = [$\xi$, $\xi$, 4] (Γ−K) respectively, while Arc branches were measured in the (400) (LEO) and (220) (LA) zones along scattering wavevectors **Q** = [4+$\xi$, 0, 0] and **Q** = [2+$\xi$, 2, 0] (Γ−M), and **Q** = [4+$\xi$, $\xi$, 0] and **Q** = [2+$\xi$, 2, 0] (Γ−K) respectively. The peak intensity at each *q* point was rescaled.

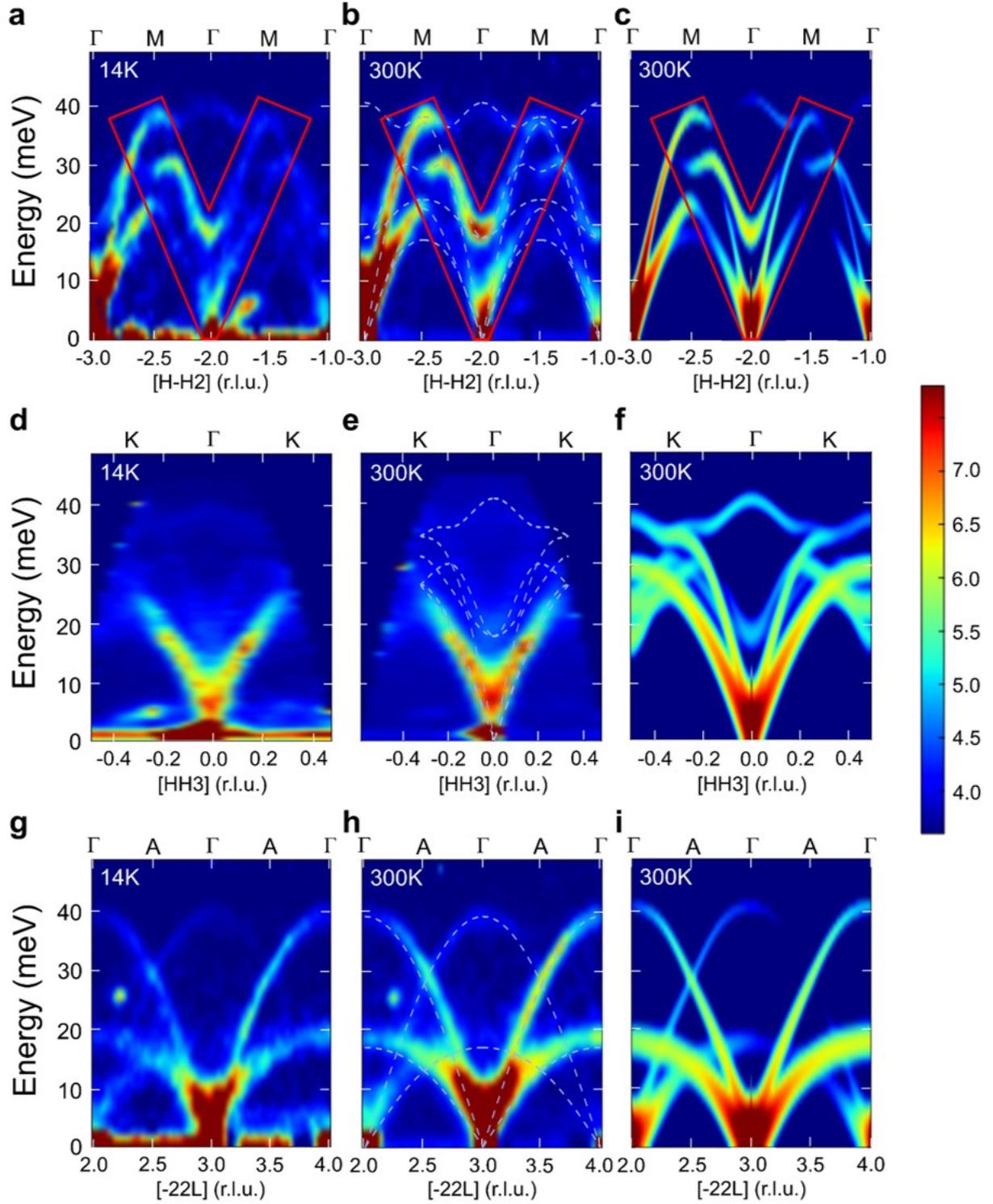

**Fig. 2. Phonon dispersion nesting observed in the $\chi''(\mathbf{Q}, E)$ from INS measurements.** (**a**), (**b**) and (**c**) are the $\chi''(\mathbf{Q}, E)$ for **Q** along [H, -H, 2] (r.l.u.). (**d**), (**e**) and (**f**) are the results along [H, H, 3]. (**a**) and (**d**) are the results at 14 K, (**b**) and (**e**) are the results at 300 K overlaid with the calculated phonon dispersion, and (**c**) and (**f**) are the calculated $\chi''(\mathbf{Q}, E)$ at 300 K. The parallel sections were marked by the red boxes. (**g**), (**h**) and (**i**) are the results along [-2, 2, L]. Small dots ((-2, 2, 2) to (-1.5, 1.5, 2) along Γ–M and around (-2, 2, 2.2) along Γ–A) in the measured $\chi''(\mathbf{Q}, E)$ are multiple scattering artifacts. Intensities are plotted on a logarithmic scale.

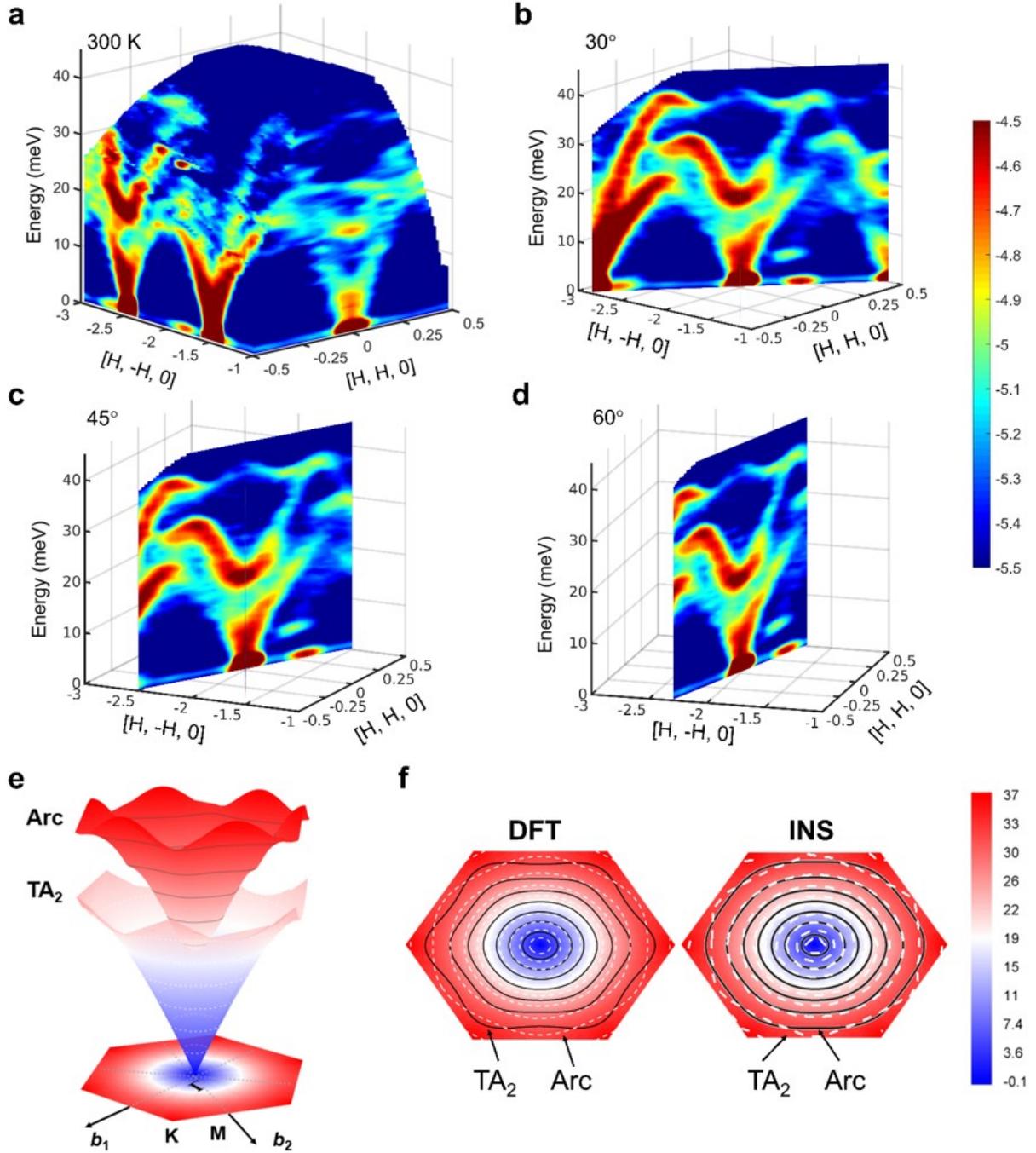

**Fig. 3. The in-plane Matryoshka-like dispersion behavior shown by INS and first-principles calculations.** (**a**), Volumetric view of the measured $\chi''(\mathbf{Q}, E)$ throughout the basal plane in the Brillouin zone at 300 K. (**b**) (**c**) and (**d**) Cuts at 30°, 45°, and 60° from the [H -H 0] direction respectively. Color bar indicates the intensities plotted on a logarithmic scale. (**e**) Side view of the $TA_2$ and Arc branches from first-principles calculation, illustrating the huge parallel region between the two surfaces. The bottom panel indicates the projection on the basal plane. (**f**) and (**g**) are the projection contours of the $TA_2$ and Arc branches on the basal plane from DFT and INS results, respectively. The white dash line and the black solid line indicate the Arc and $TA_2$ phonon surfaces' energy contours, respectively. The energy difference between the contours of the Arc and $TA_2$ phonon surfaces is 2.5 meV. The color scale indicates the phonon surface energies.

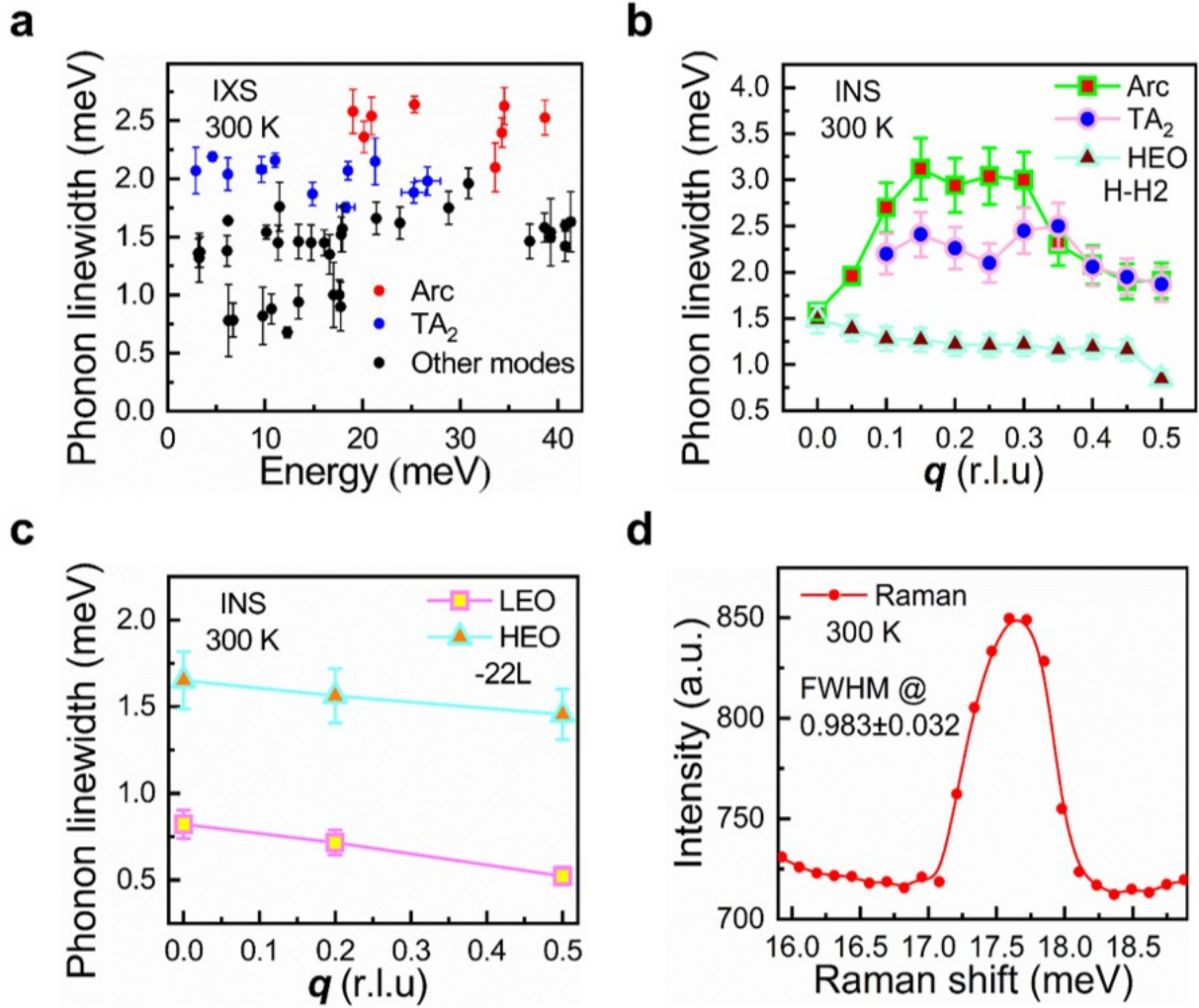

**Fig. 4. Phonon anharmonicity of the Arc branch.** (**a**) Linewidths of phonon modes from the IXS data at 300 K. (**b**) and (**c**) Phonon linewidths of the Arc, TA$_2$, and the HEO modes along Γ–M and phonon linewidths of the HEO and LEO modes along Γ–A from the INS data at 300 K. (**d**) Phonon mode and the linewidth of the Arc at Γ point from Raman data at 300 K. Raman spectroscopy only provides information on Raman-active modes at zone centers (Γ point). The error bars are from fitting uncertainties.

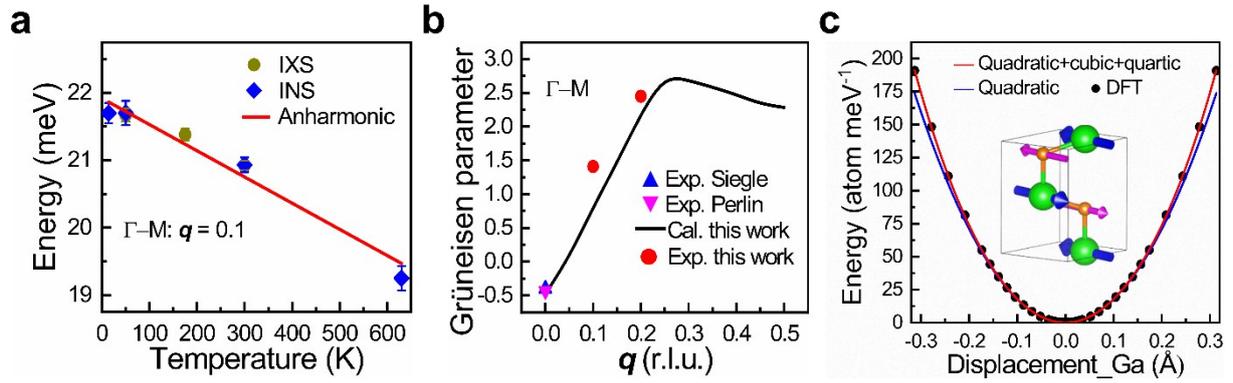

**Fig. 5. Anharmonicity of the optical mode in the Arc at $q = 0.1$ along $\Gamma-M$ in $\alpha$-GaN.** (**a**) Temperature dependence of phonon energy. Data were obtained from both IXS and INS measurements. The lines show the fitting of the data points by equation (2). The error bars are from fitting uncertainties. (**b**) The experimental mode Grüneisen parameters of the phonon modes on the Arc branch, compared with the DFT results along $\Gamma-M$. The blue triangle and the pink inverted triangle are the data from the Raman measurements reported by Siegle et al [51] and Perlin et al [52] respectively. (**c**) The frozen phonon potential of the Arc at $q = 0.1$ along $\Gamma-M$ corresponding to the displacement of Ga along $b$. Inset indicates the corresponding vibration mode, where the green balls represent Ga atoms and orange balls represent N atoms. The vectors indicate the direction of the displacement of the corresponding atoms.

**Table 1 | Group velocity of acoustic phonons along the high symmetry directions from experimental data. Relaxed DFT results are listed for comparison.**

| Direction | Temperature (K) | LA $v_g$ (m/s) | TA$_2$ $v_g$ (m/s) |
|---|---|---|---|
| Γ–M [100] | 14 | 8319 | 4142 |
|  | 50 | 8149 | 4008 |
|  | 300 | 8015 | 3894 |
|  | DFT | 7909 | 4088 |
| Γ–K [110] | 50 | 7521 | 3768 |
|  | 300 | 7363 | 3466 |
|  | DFT | 7465 | 3941 |
| Γ–A [001] | 14 | 8403 | 4303 |
|  | 50 | 8290 | 4135 |
|  | 300 | 8087 | 4062 |
|  | DFT | 7987 | 3984 |

# Matryoshka Phonon Twinning in $\alpha$-GaN

Bin Wei[1,2,3], Qingan Cai[2], Qiyang Sun[2], Yaokun Su[4], Ayman H. Said[5], Douglas L. Abernathy[6], Jiawang Hong[1]*, and Chen Li[2,4]*

[1]School of Aerospace Engineering, Beijing Institute of Technology, Beijing 100081, China
[2]Department of Mechanical Engineering, University of California, Riverside, Riverside, CA 92521, USA
[3]Henan Key Laboratory of Materials on Deep-Earth Engineering, School of Materials Science and Engineering, Henan Polytechnic University, Jiaozuo 454000, China
[4]Materials Science and Engineering, University of California, Riverside, Riverside, CA 92521, USA
[5]Advanced Photon Source, Argonne National Laboratory, Lemont, IL 60439, USA
[6]Neutron Scattering Division, Oak Ridge National Laboratory, Oak Ridge, TN 37831, USA

*e-mail: chenli@ucr.edu; hongjw@bit.edu.cn

# 1. Structure and sample information of $\alpha$-GaN.

Fig. S1 shows the crystal structure of $\alpha$-GaN and the single crystal sample information. GaN adopts the wurtzite ($\alpha$) structure (hexagonal) under ambient conditions, composed of two interpenetrating hcp lattices of Ga and N atoms, with one Ga atom and four neighboring N atoms forming a tetrahedral structure (Fig. S1a). Fig. S1b shows the Brillouin Zone of $\alpha$-GaN. Fig. S1c shows the single crystal samples for the INS (upper) and IXS (lower) measurements, respectively. Fig. S1d shows the rocking curve of our sample, which indicates high crystalline quality.

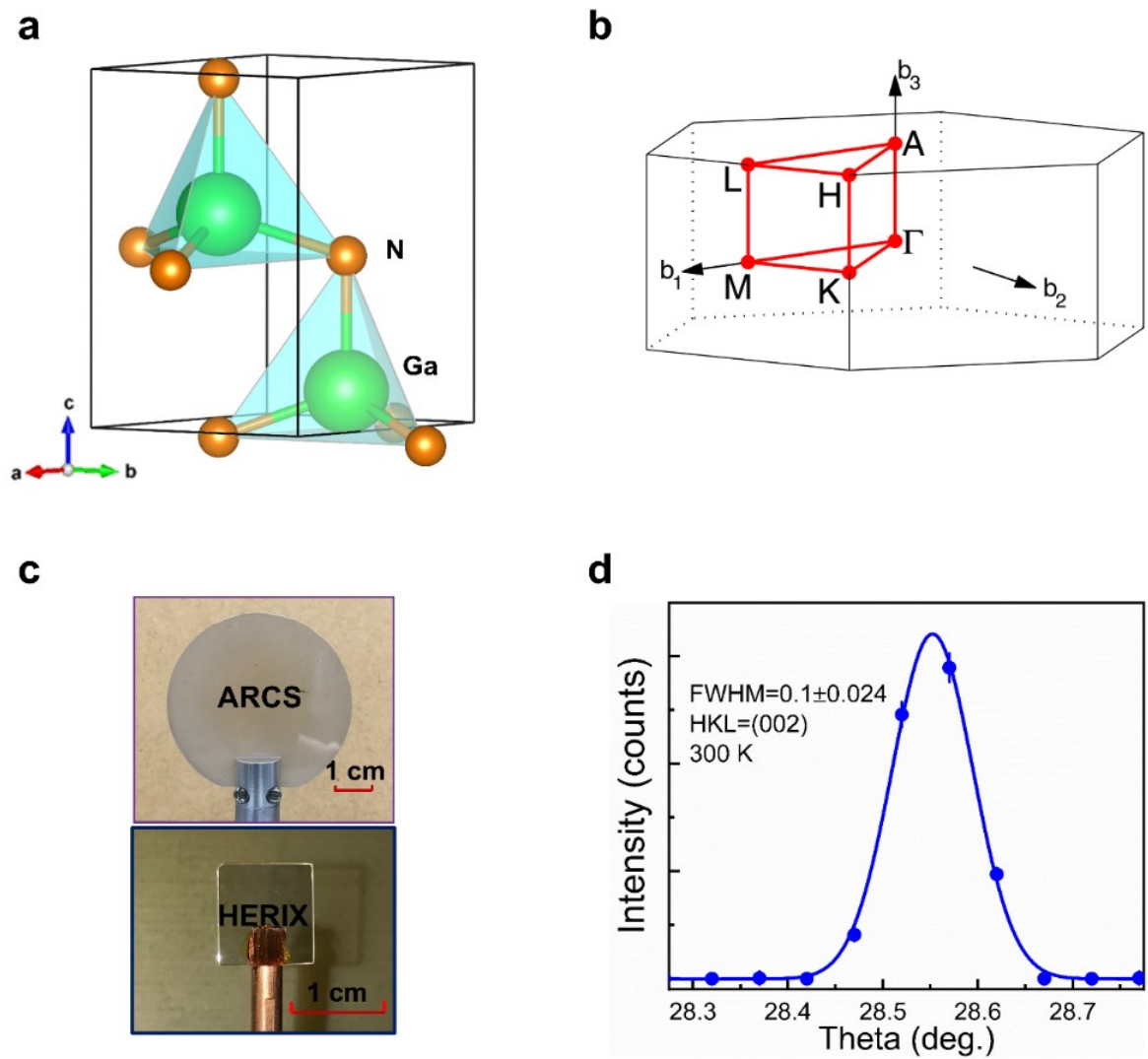

**Fig. S1. Crystal structure of $\alpha$-GaN and single crystal sample information.** (**a**) Crystal structure of $\alpha$-GaN. (**b**) The first BZ of the $\alpha$-GaN hexagonal lattice. (**c**) A single-crystal substrate with a diameter of 5 cm mounted for INS measurement (upper) and a small square one with a side of 1 cm for IXS measurement (lower). (**d**) The rocking curve of the $\alpha$-GaN single crystal for (002) Bragg peak by X-ray showing the high quality of the sample.

## 2. Schematic of Matryoshka phonon dispersion twinning in α-GaN.

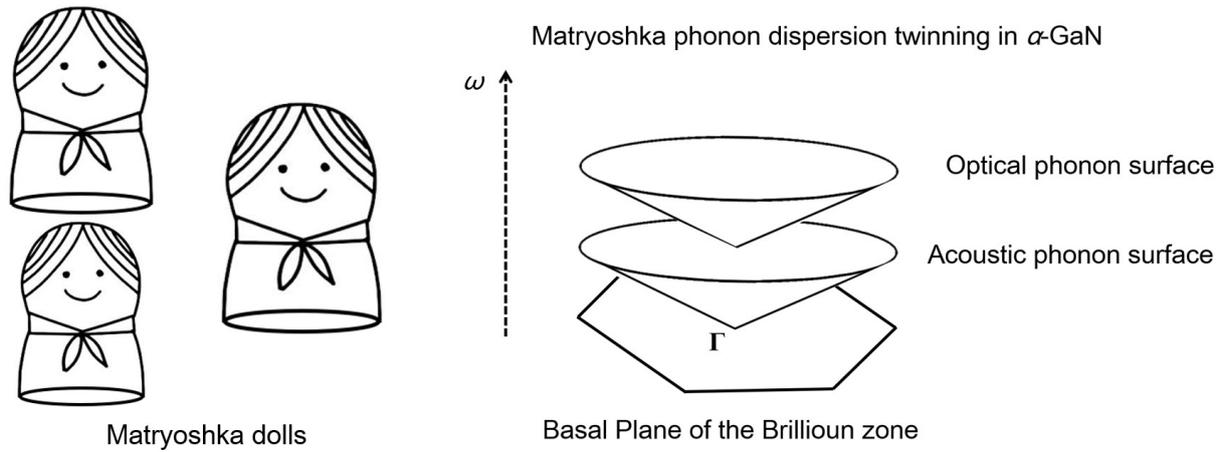

**Fig. S2. Schematic of Matryoshka phonon dispersion twinning in α-GaN.** Matryoshka dolls (left). A set of dolls of decreasing size placed one inside another. Matryoshka phonon dispersion twinning in α-GaN (right).

## 3. Inelastic X-ray scattering measurements of α-GaN.

Fig. S3 shows the phonon dispersion of α-GaN measured by IXS at 50 and 175 K below the gap around 45 meV. The two parallel sections were also found along Γ–M and Γ–K (see stars labeled branches), showing the dispersion nesting behavior.

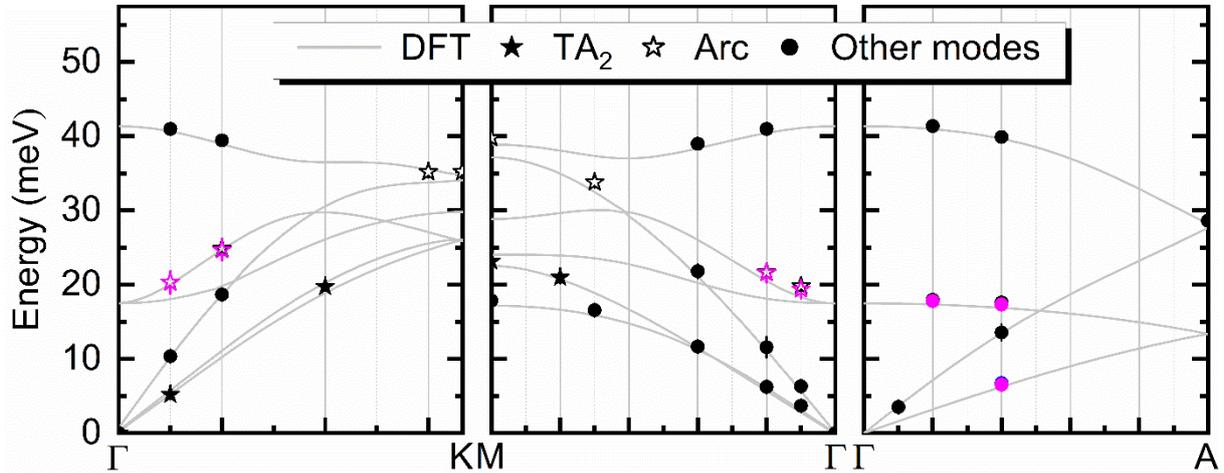

**Fig. S3. Phonon dispersion nesting observed by IXS measurements at 50 K (black) and 175 K (magenta), overlaying on the calculation (grey line).** Along Γ–M and Γ–K, the $TA_2$ (solid stars) and the Arc (hollow stars) phonon jointly shows the nesting behavior. The error bars are from fitting uncertainties. Most errors are comparable to or smaller than the data point symbols.

Fig. S4 shows how the measured longitudinal acoustic (LA) and high energy optic (HEO) phonon peaks evolve with momentum transfer at 300 K. It can be observed that the peak centers (phonon energies) of the LA modes exhibit a dispersive behavior with increasing $q$, while the HEO modes exhibit a non-dispersive behavior.

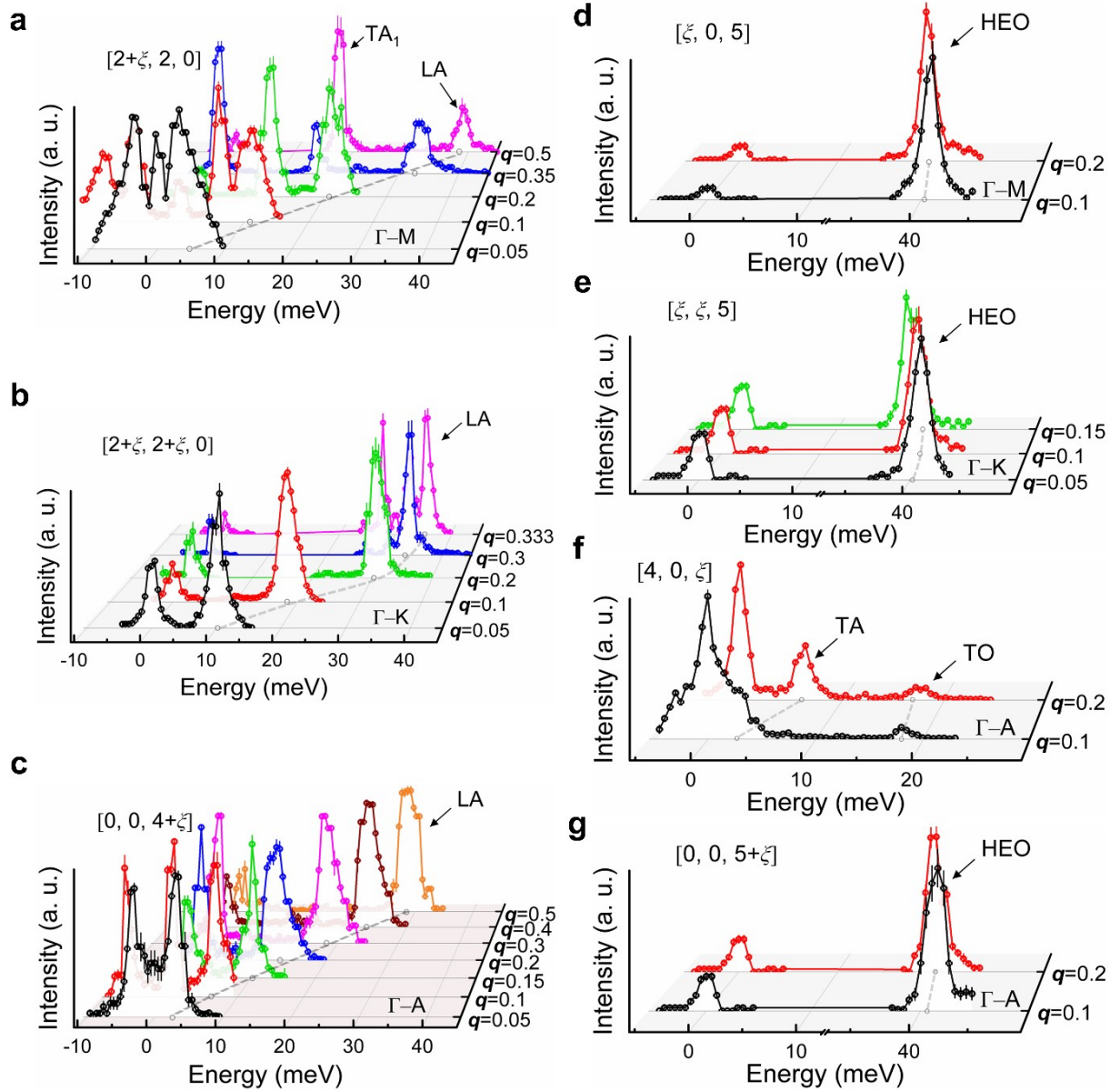

**Fig. S4. The measured phonon spectra at several $q$'s at 300 K.** The phonon dispersion is illustrated by the grey dash line. The peak centers are labeled by circles. LA branch is measured in the (220) zone along $\mathbf{Q} = [2+\xi, 2, 0]$ (Γ–M) (**a**), along $\mathbf{Q} = [2+\xi, 2+\xi, 0]$ (Γ–K) (**b**), and measured in the (004) zone along $\mathbf{Q} = [0, 0, 4+\xi]$ (Γ–A) (**c**). HEO branches are measured in the (005) zone along scattering wavevector $\mathbf{Q} = [\xi, 0, 5]$ (Γ–M) (**d**), $\mathbf{Q} = [\xi, \xi, 5]$ (Γ–K) (**e**), and $\mathbf{Q} = [0, 0, 5+\xi]$ (Γ–A) (**g**), respectively. (**f**) TA and TO branches are measured in the (400) zone along $\mathbf{Q} = [4, 0, \xi]$ (Γ–A). The peak intensity at each $q$ point is rescaled.

Fig. S5 shows the constant-**Q** IXS spectra for some optical modes of $\alpha$-GaN at 50, 175, and 300 K. It can be observed that phonon modes soften and broaden with increasing temperature.

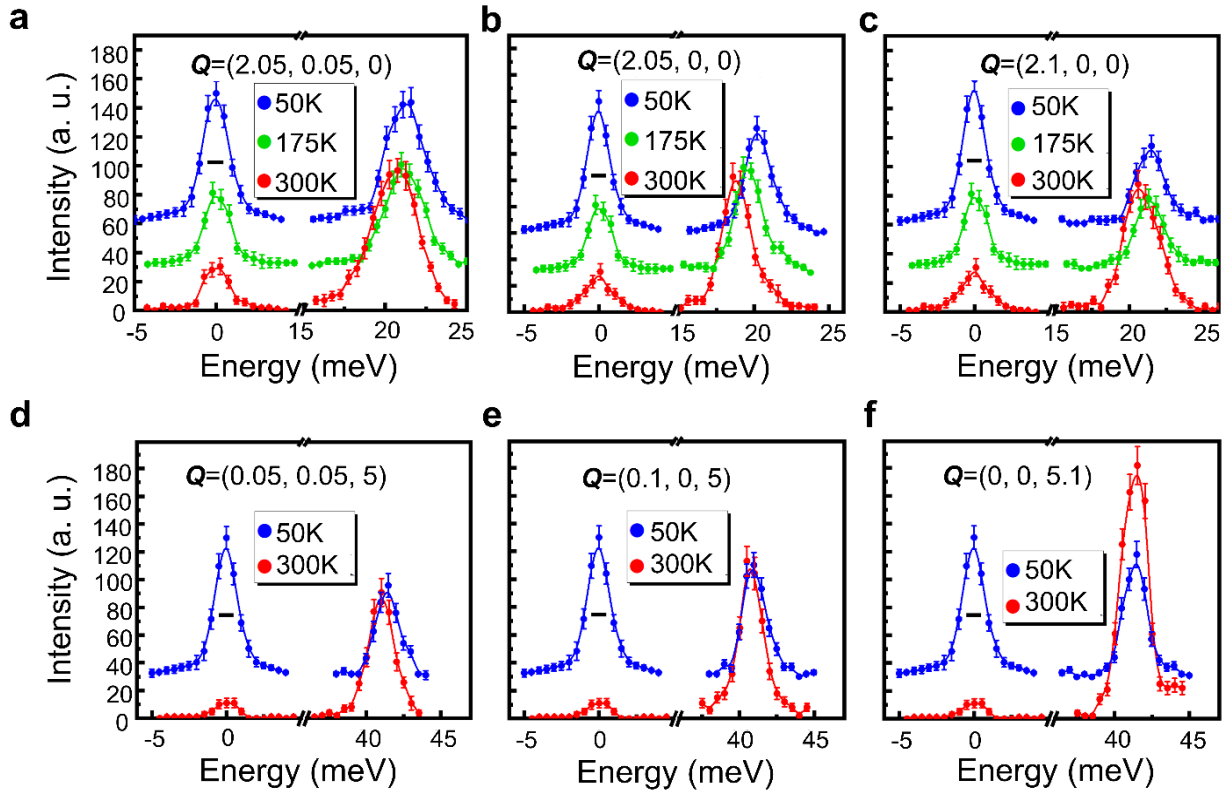

**Fig. S5. Constant-Q IXS spectra for the nested optical modes of $\alpha$-GaN at 50, 175, and 300 K.** The left peak is the elastic peak, and the right peak is the phonon mode in each spectrum. The curves are DHO fitting, and the short black lines at the elastic peak indicate the instrument resolution. Error bars are from counting statistics. **a-c**, Arc modes are at **Q** = (2.05, 0.05, 0), (2.05, 0, 0), and (2.1, 0, 0), respectively, at 50, 175 and 300 K. **d-f**, The HEO mode are at **Q** = (0.05, 0.05, 5), (0.1, 0, 5), and (0, 0, 5.1), respectively, at 50 and 300 K.

**4. Inelastic neutron scattering measurements of $\alpha$-GaN.**

Figures S6a-S6d plot the volumetric and the internal views of the measured $\chi''(\mathbf{Q}, E)$ throughout the basal plane in the Brillouin zone of $\alpha$-GaN at 14 K, which provide a visualization of this nesting behavior between the TA$_2$ and the Arc branches. Figures S6b-S6d are cuts from Fig. S6a at 30°, 45°, and 60° from [H -H 0], respectively. These phonons at non-high-symmetry directions demonstrate that the nesting behavior is universal throughout the basal plane.

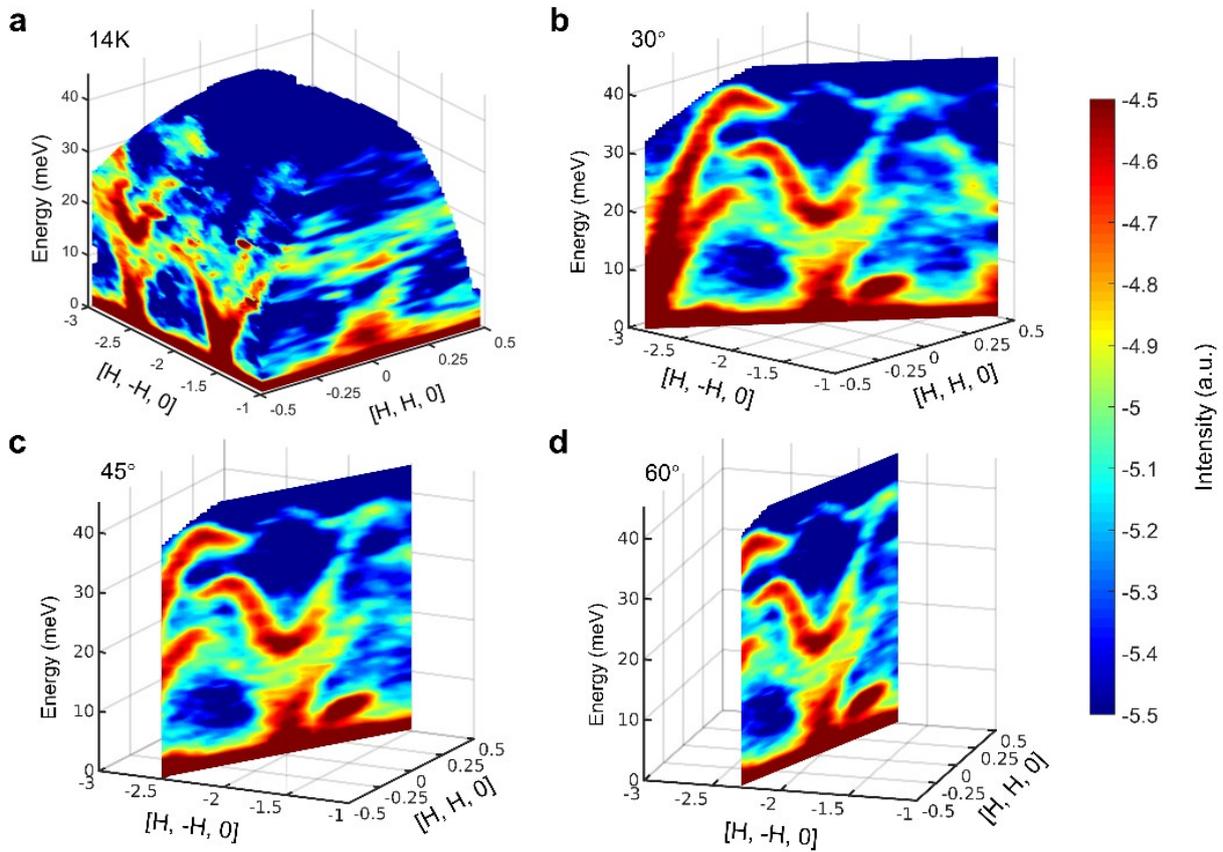

**Fig. S6. The in-plane Matryoshka-like dispersion behavior shown by INS at 14 K.** (**a**) Volumetric view of the measured $\chi''(\mathbf{Q}, E)$ throughout the basal plane in the Brillouin zone. (**b**), (**c**), and (**d**) Cuts at 30°, 45°, and 60° from [H -H 0], respectively. Color bar indicates the intensities plotted on a logarithmic scale.

Fig. S7 shows additional views of the cuts from Fig. 3a (300 K) and Fig. S6 (14 K) at 30°, 45°, and 60° from [H -H 0] direction, respectively.

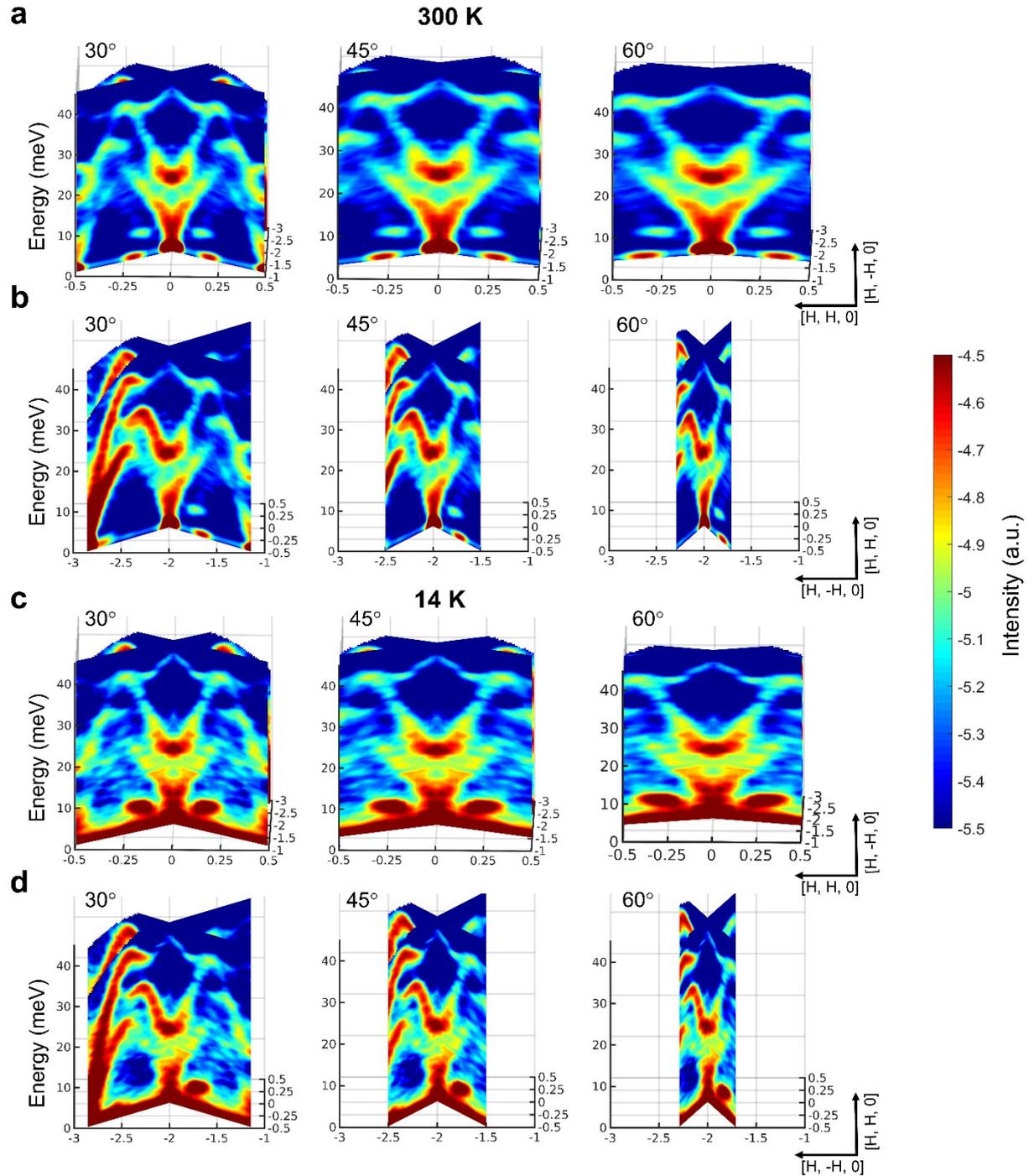

**Fig. S7. Cuts from the INS results at 30°, 45°, and 60° from the [H -H 0] direction at both 14 and 300 K.** (**a**) and (**b**) are the view directions along [H H 0] and [H -H 0] at 300 K, respectively. (**c**) and (**d**) are the results at 14 K. Color bar indicates the intensities plotted on a logarithmic scale.

## 5. Matryoshka-like behavior confirmed by both the first-principles calculations and INS.

Fig. S8a shows the calculated phonon dispersion of $\alpha$-GaN. The blue shadows clearly reflect the near-constant energy differences on the nested $TA_2$ and Arc branches along $\Gamma$–M, $\Gamma$–T, $\Gamma$–S, and $\Gamma$–K directions, leading to the Matryoshka dispersion twinning behavior. Besides, the high energy optical phonons in $\alpha$-GaN are mostly non-dispersive and have a large energy gap of about 20 meV above the low energy phonons, which indicates that the high-energy phonons will not significantly scatter the acoustic phonons [1]. Fig. S8b shows the projection contours of the $TA_2$ and Arc branches on the basal plane extracted from the first-principles calculation (DFT) and INS results. It can be seen that the energy difference between the $TA_2$ and Arc contours is 2.5 meV, and the contours keep the circular shape even up to the high energy region, especially for the INS results. This observation indicates that the shape of the Matryoshka twinning keeps well throughout the basal plane.

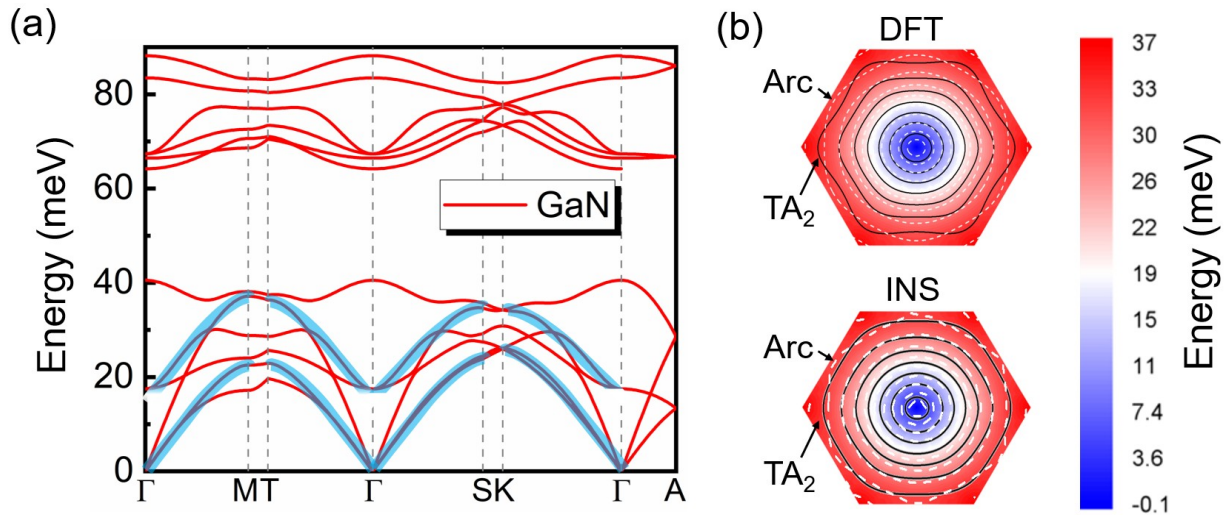

**Fig. S8. The in-plane Matryoshka-like dispersion behavior shown by INS and first-principles calculations.** (**a**) The blue shadowed areas represent the phonon nesting modes throughout the basal plane of the Brillouin zone along $\Gamma$–M, $\Gamma$–T, $\Gamma$–S, and $\Gamma$–K directions. M (1/2, 0, 0) and K (1/3, 1/3, 0) are the high symmetry points of the Brillouin zone, and T (4/9, 1/9, 0) and S (7/18, 2/9, 0) are points on the path from M to K. (**b**) The projection contours of the $TA_2$ and Arc branches on the basal plane from the first-principles calculation (DFT) and INS results. The white dash line and the black solid line indicate the $TA_2$ and Arc phonon surfaces' energy contours, respectively. The difference between the contours of the $TA_2$ and Arc phonon surfaces is 2.5 meV. The color scale indicates the phonon surface energies.

## 6. Schematic of the three-phonon scattering channels enhanced by the Matryoshka dispersion twinning in $\alpha$-GaN.

The twinning of the Matryoshka-like dispersion potentially provides a vast number of three-phonon scattering channels in $\alpha$-GaN by reducing the momentum transfer constraint. For example, in the three-phonon emission process where any phonon mode of the Arc branch ($q_{Arc}$, $\omega_{Arc}$) decays into two phonon modes, for one mode of the Arc branch around $\Gamma$ point ($q_{Arc0}$, $\omega_{Arc0}$), we can always find a TA$_2$ mode ($q_{Arc} - q_{Arc0}$, $\omega_{Arc} - \omega_{Arc0}$) in the basal plane that enables such the scattering (see Fig. S9). In contrast, for phonon modes at higher or lower energy, the number of emission channels will be less, and the heat carrier (mainly acoustic phonon) involved will be scattered less strongly.

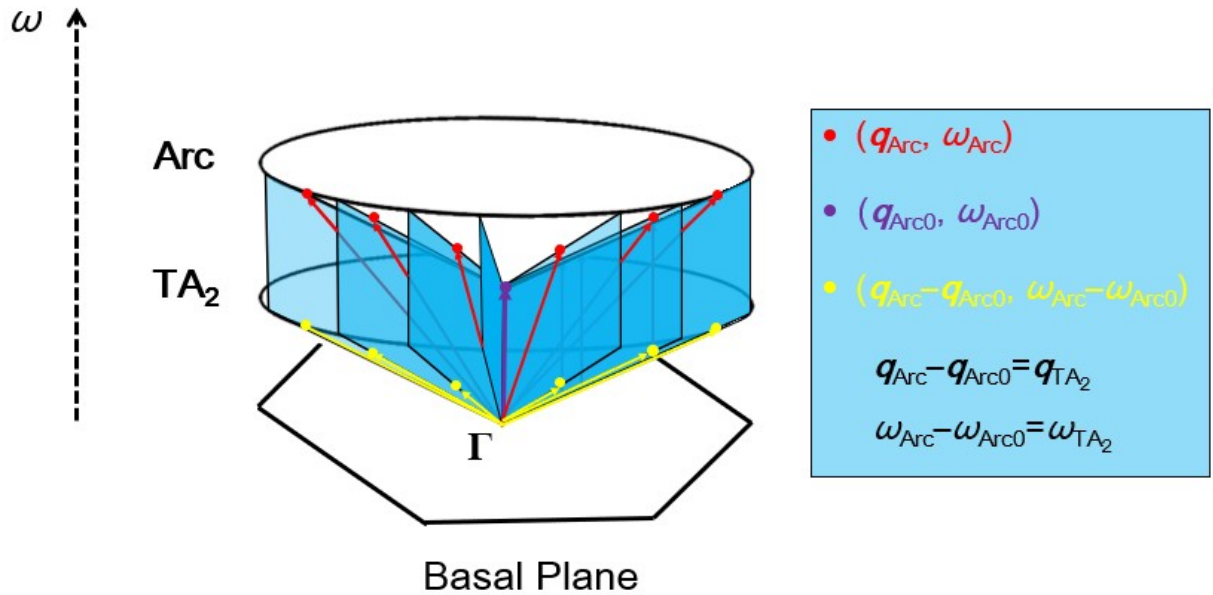

**Fig. S9. Schematic of the three-phonon scattering channels enhanced by the Matryoshka dispersion twinning in $\alpha$-GaN.** The two cones represent the Arc and the TA$_2$ phonon energy surfaces. The red, purple, and yellow dots represent the phonon modes on the Arc phonon surface, the phonon mode around $\Gamma$ point of the Arc surface (conic top), and the phonon modes on the TA$_2$ phonon surfaces.

## 7. Calculated weighted phase space of phonon scattering in α-GaN.

Fig. S10 shows the weighted phase space of three-phonon scattering in α-GaN, which is reproduced from ref. [2] under the author's permission. The weighted phase space below 40 meV is much larger than that above 70 meV. Moreover, the values of the emission process are larger than those of the absorption process, especially in the energy range from 20 to 40 meV. Such behavior is consistent with the reported Matryoshka phonon twinning.

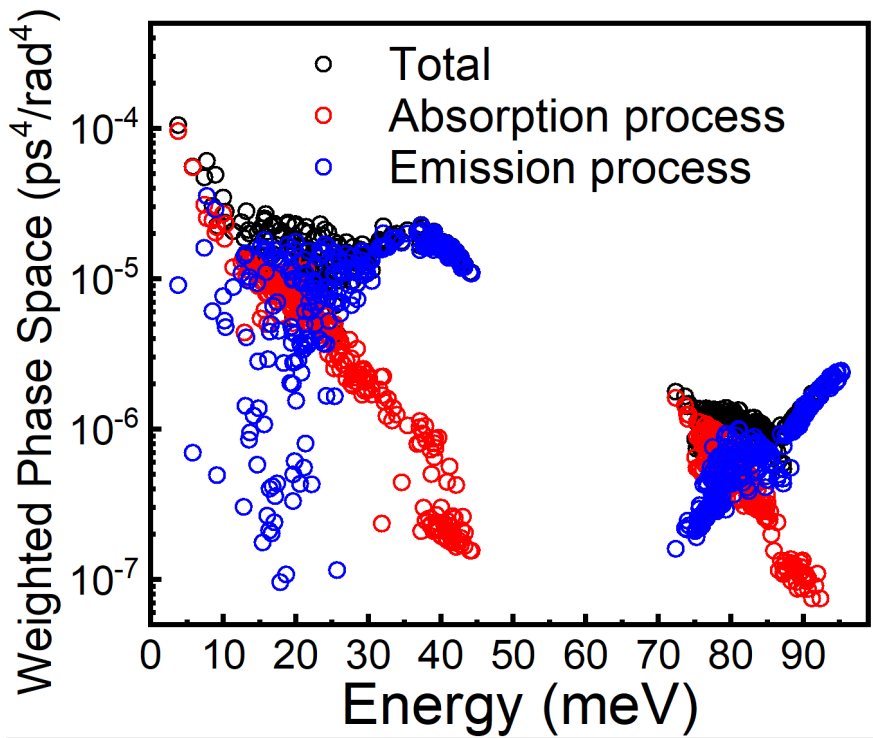

**Fig. S10. Calculated weighted phase space as a function of phonon energy in α-GaN [2].** The black circles indicate the total weighted phase space, the red circles indicate the absorption process, and the blue circles indicate the emission process.

## 8. Grüneisen parameter calculations.

Grüneisen parameters describe how much the phonon frequency shifts with changes in volume. The mode Grüneisen parameter is defined as [3]

$$\gamma_j(q) = -\frac{V_0}{\omega_j(q)}\frac{\partial \omega_j(q)}{\partial V} \qquad (1)$$

where $V_0$ is the equilibrium volume, $\omega_j(\mathbf{q})$ is the phonon energy of wavevector $\mathbf{q}$ and branch index $j$. Phonon dispersions calculated with the volumes 1% larger and smaller than the equilibrium volume are used to extract the quasi-harmonic Grüneisen parameters, as implement in Phonopy code [4].

The experimental mode Grüneisen parameter was obtained based on the isobaric Grüneisen parameter $\gamma_P(T)$ concept, which is usually used to quantify the temperature dependence of the phonon energy shifts and is defined as [5]

$$\gamma_P(T) \equiv -\frac{1}{3\alpha(T)}\left\langle\frac{\partial \ln \omega_i}{\partial T}\right\rangle\bigg|_P \qquad (2)$$

where $\alpha$ is the thermal expansion coefficient of $\alpha$-GaN [6], and $\omega_i$ is the phonon mode energy. The difference between quasi-harmonic and isobaric Grüneisen parameters is a good indication of phonon anharmonicity [7]. In Fig. 5b, the mode Grüneisen parameters at 300 K show only a minor discrepancy between the two. This result implies that the measured isobaric Grüneisen parameters of $\alpha$-GaN contain contributions mostly from the quasi-harmonic model instead of the anharmonicity [8].